\documentclass[twoside]{article}
\usepackage{epstopdf,amsfonts,latexsym}
\usepackage{euscript,amssymb,epsfig}
\usepackage{graphicx}
\usepackage{epstopdf}
\usepackage{natbib}
\usepackage{tabularx}

\RequirePackage{color}

\usepackage{aat}

\newcommand{\be}[1]{\begin{equation}\label{#1}}
\newcommand{\ee}{\end{equation}}
\newcommand{\ba}[1]{\begin{eqnarray}\label{#1}}
\newcommand{\ea}{\end{eqnarray}}
\newcommand{\rf}[1]{(\ref{#1})}
\newcommand{\nn}{\nonumber}

\begin{document}
\thispagestyle{myfirst}

\setcounter{page}{1}

\vspace*{2.5cm}
\mytitle{Dark and visible matter distribution in Coma cluster: theory vs observations}
\myauthor{R.D. Brilenkov$^{1,2,}$\footnote{e-mail: ruslan.brilenkov@gmail.com}, M.V. Eingorn$^3$, A.I. Zhuk$^4$}
\myadress{$^1$ Institute for Astro- and Particle Physics, University of Innsbruck,\\ 25/8 Technikerstrasse, Innsbruck A-6020, Austria \\
$^2$ Dipartimento di Fisica e Astronomia `G. Galilei', Universit\`{a} di Padova,\\ 3 vicolo dell'Osservatorio, Padova 35122, Italy \\
$^3$ North Carolina Central University, CREST and NASA Research Centers,\\ 1801 Fayetteville Street, Durham, North Carolina 27707, U.S.A.\\
$^4$ Astronomical Observatory, Odessa National University,\\ 2 Dvoryanskaya Street, Odessa 65082, Ukraine\\ }

\mydate{Received 15 February, 2016}

\myabstract{We investigate dark and visible matter distribution in the Coma cluster in the case of the Navarro-Frenk-White (NFW) profile. A toy model where all
galaxies in the cluster are concentrated inside a sphere of an effective radius $R_{\mathrm{eff}}$ is considered. It enables us to obtain the mean velocity
dispersion as a function of $R_{\mathrm{eff}}$. We show that, within the observation accuracy of the NFW parameters, the calculated value of $R_{\mathrm{eff}}$
can be rather close to the observable cutoff of the galaxy distribution. Moreover, the comparison of our toy model with the observable data and simulations leads
to the following preferable NFW parameters for the Coma cluster: $R_{200} \approx 1.77\,h^{-1} \, \mathrm{Mpc} = 2.61\, \mathrm{Mpc}$, $c=3\div 4$ and $M_{200}=
1.29h^{-1}\times10^{15}M_{\odot}$. In the Coma cluster the most of galaxies are concentrated inside a sphere of the effective radius $R_{\mathrm{eff}}\sim 3.7$
Mpc and the line-of-sight velocity dispersion is $1004\, \mathrm{km}\, \mathrm{s}^{-1}$.} \mykey{Cosmology: dark matter, Galaxies: clusters: general, Galaxies:
clusters: individual: Coma, Galaxy: halo, Haloes: density profiles}

\section{\label{sec:1} Introduction}

According to the recent observations \citep{Ries,Perlmutter,Eisenstein,Hinshaw,Planck,Planck2}, our Universe is dark: dark energy and dark matter contribute
approximately 69\% and 26\% into total mass-energy balance in the Universe, respectively. Different independent observations also indicate that dark matter (DM)
envelops the galaxies and clusters of galaxies. Baryonic matter is only 10-15\% of the total mass of clusters of galaxies \citep{clustermass}. The rest is DM.
Unfortunately, the nature of DM (as well as dark energy) is still unclear and is a subject of numerous investigations \citep{ourbook}. The standard cosmological
model, i.e. the $\Lambda$CDM model, assumes that DM is cold. This model  is rather successful in explaining the structure formation in the Universe. There is a
large number of papers devoted to the investigations (analytical and numerical) of the structure and dynamics of galaxies and clusters of galaxies (see, e.g., the
reviews \citep{minireview1} and \citep{minireview2}). For these investigations, the spherical density profile $\rho(R)$ of DM in galaxies and clusters of galaxies
is very useful. On the one hand, such profiles help to reveal different universal scaling relations \citep{Boyar1,Boyar2,selfsim}. On the other hand, they help to
study the observable properties of DM in haloes of galaxies and clusters of galaxies \citep{clustermass,Okabe,SabPop} and are used in the recent numerical
simulations \citep{Sch}. These profiles also represent a very useful tool for modeling the mass distribution in haloes \citep{Chemin} as well as for analytical
analysis of properties of DM haloes \citep{Chernin}.

Galaxies, groups and clusters of galaxies are the inhomogeneities we observe inside the cell of uniformity which is of the order of 190 Mpc \citep{EZcosm2} (for
the estimation of the homogeneity scale from the point of view of gravitational interaction features at large distances see also \citep{XXX}). Deep inside this
cell, the Universe is highly inhomogeneous and is well described by the discrete cosmology approach \citep{EZcosm1,EZcosm2,XXX}. The structure and dynamics of
inhomogeneities are influenced by two main factors. They are the gravitational attraction between the constituents of these objects and the cosmological expansion
of the Universe. Obviously, there is a distance from the center of mass of an inhomogeneity at which the cosmological expansion begins to prevail over the
gravitational attraction. Moreover, because the expansion in the late Universe is accelerating, both of these phenomena act against each other. This means that
the corresponding forces are directed oppositely. This effect was observed experimentally \citep{Kar2002,Kar2012,Kar2008,KarNas} and the corresponding distance
was called the radius of the zero velocity surface. In \citep{Chernin1} this distance has received the name of the radius of zero gravity. In our opinion, this
distance is more properly termed as the radius of the zero acceleration \citep{EZcosm1,EZcosm2} because gravity does not disappear at this distance but the
acceleration of a test body (a dwarf galaxy) is equal (or approximately equal) to zero. This distance for the Local Group was estimated in \citep{EKZ2}.

It is natural to suppose that the edge of the DM halo in clusters of galaxies corresponds to the surface of the zero acceleration. Therefore, the radius of this
surface can be considered as the size of clusters. This idea was discussed in \citep{Chernin}. Here, the authors considered the background metric in the form of
the Schwarzschild-de Sitter (SdS) solution. This approach disregards the presence of matter in the Universe (the alternative reasoning is available in
\citep{IJMPD}). However, in the $\Lambda$CDM model matter contributes 31\% into the total mass-energy balance. Therefore, in this paper we take matter into
account considering the Friedmann-Robertson-Walker (FRW) metric as the background metric.

In Section 2, we start from the investigation of the gravitational potential produced by a spherically distributed DM halo. This potential satisfies the Poisson
equation \citep{EZcosm1}. We supplement this equation with the proper boundary conditions at the surface of zero acceleration. Then, with the help of the virial
theorem, we define the mean velocity dispersion of the galaxies which we consider as test bodies in the cluster DM halo. Here we use the observation \citep{Tully}
that  distribution of galaxies in rich clusters is abruptly bounded at some distances much smaller than the size of a cluster. Thus, we introduce an effective
radius $R_{\mathrm{eff}}$ and suppose that all galaxies are concentrated inside a sphere of this radius. In Section 3, we apply the derived formulae to the case
of the Navarro-Frenk-White (NFW) DM profile for the Coma cluster. Here, first, we demonstrate that $R_{\mathrm{eff}}$ is close to the observable value if we use
the NFW profile parameters for the Coma cluster found from the observations and simulations. Second, the comparison of our toy model with the observable data and
simulations enables us to get preferable NFW profile parameters for the Coma cluster. The main results are briefly summarized in concluding Section 4.

\section{\label{sec:2} Zero acceleration sphere and mean velocity dispersion}

In the framework of discrete cosmology developed in the recent series of papers \citep{EZcosm1,EZcosm2,EKZ2,XXX} the acceleration of any test body in the
gravitational field of a spherically symmetric
%{\footnote{This restriction does not affect the main conclusions of our paper.}}
gravitationally bound system is given by the formula (see, e.g., section 4 in \citep{EZcosm1})
%%%%%%
\be{2.1}
\ddot R = \frac{\ddot a}{a}R-\frac{\partial \Phi}{\partial R}\, ,
\ee
%%%%%%
where $R$ is the distance from the center of mass of the system to a test body, $a$ is the scale factor of the Universe, the overdots denote the derivatives with
respect to the synchronous time, and the gravitational potential $\Phi$ satisfies the Poisson equation
%%%%%%
\be{2.2} \triangle\Phi=\frac{1}{R}\frac{d^2}{dR^2}(R\Phi) = 4\pi G_N\rho_{\mathrm{ph}}\, , \ee
%%%%%%
where $G_N$ is the Newtonian gravitational constant and $\rho_{\mathrm{ph}}$ is the physical (not comoving!) rest mass density of the bound system. It is worth
noting that Eq. \rf{2.1} was obtained in \citep{EZcosm1} for the Universe with hyperbolic spatial topology $\mathcal{K}=-1$. However, deep inside the cell of
uniformity (with the size of the order of 190 Mpc \citep{EZcosm2}) the Universe can be considered with very high accuracy as spatially flat. Therefore, we can
drop the contribution of $\mathcal{K}$ to \rf{2.2} and consider the Laplace operator $\triangle$ in cartesian (physical) coordinates. It should be mentioned that
in the Newtonian limit the similar equations (of the form \rf{2.1} and \rf{2.2}) were obtained in \citep{Peebles} and are used for the N-body simulations
\citep{gadget2}. It is important to emphasize as well that Newtonian gravity is suitable for subhorizon scales, while for covering the whole space one needs to
resort to the relativistic perturbation theory \citep{XXX} (see also the corresponding generalizations to the second order \citep{BrilEin} and various fluids in
the Universe \citep{Brilenkov,Kiefer}).

The term $(\ddot a/a)R$ in \rf{2.1} arises due to the global cosmological expansion of the Universe while the term with $\Phi(R)$ takes into account the
gravitational attraction between a test body and the gravitationally bound system. According to recent observations, our present Universe undergoes the stage of
the accelerated expansion. For the $\Lambda$CDM model, this stage started at the redshift $z\approx 0.76$ \citep{accel}. It means that from this time $\ddot a>0$.
Therefore, Eq. \rf{2.1} demonstrates that (starting from this time) there is a distance $R_H$ from the center of mass of the system at which the gravitational
attraction and the cosmological expansion compensate each other:
%%%%%
\be{2.3}
\left.\ddot R\right|_{R_H}=0\quad \Rightarrow \quad R_H=\left.\frac{\partial\Phi/\partial R}{\ddot a/a}\right|_{R_H}\, .
\ee
%%%%%
Obviously, the Hubble flows begin to
%prevail over the gravitational attraction
form at distances greater than $R_H$.  We can also rewrite Eq. \rf{2.1} in the form
%%%%%
\be{2.4}
\ddot R = - \frac{\partial \widetilde \Phi}{\partial R}\, ,
\ee
%%%%%
where
%%%%%
\be{2.5}
\widetilde\Phi(R)=-\frac{1}{2}\frac{\ddot a}{a}R^2 + \Phi(R)\, .
\ee
%%%%%
Then, the zero acceleration condition is
%%%%%
\be{2.6}
\left.\frac{\partial \widetilde\Phi}{\partial R}\right|_{R_H}=0\, .
\ee
%%%%%

In this paper, we consider gravitationally bound systems consisting of a rather large number of inhomogeneities/galaxies which have common (for a whole
system) DM haloes. DM haloes make a dominant contribution to the total mass of the systems and galaxies are considered as test bodies. This situation takes place
in rich clusters of galaxies, e.g., for the Virgo cluster which consists of about 2000 galaxies \citep{KarNas,BinSanTamm1}. It is well known that the shape of this
cluster is far from being spherically symmetric \citep{BinSanTamm2,Boselli}. The cluster is an aggregate of at least three separate subclumps. In general, there
is no problem to define a surface (or an approximate surface) of zero acceleration in the case of a number of gravitating sources which make the dominant
contributions to the gravitational acceleration  \citep{EKZ2}. However, the DM profiles for galaxies and clusters of galaxies are usually modeled by the
spherically symmetric distributions of DM. The Coma cluster (Abell 1656) has more spherical distribution\footnote{Although here we also observe subhaloes
\citep{subhalos}.}. Therefore, in what follows we consider the Coma cluster which, in a rough approximation, has an overall DM halo of the spherically symmetric
form.

It is natural to suppose that the edge/cutoff of a DM halo coincides with the surface of zero acceleration. DM particles are confined by the gravitational
interaction inside the domain bounded by this surface and are ``blown" by the cosmological expansion outside it. According to the observations, DM makes the main
contribution into the mass of galaxies, groups and clusters of galaxies. Then, the rest mass density $\rho_{\mathrm{ph}}$ in Eq. \rf{2.2} is mainly defined by DM.
For simplicity neglecting the contribution of the visible matter as compared with that of DM, for the halo mass we obtain
%%%%
\be{2.7}
M=4\pi \int\limits_{0}^{R_H}\rho_{\mathrm{ph}}(R) R^{2}dR\, .
\ee
%%%%%
It is clear that the gravitational potential $\Phi$ should satisfy the boundary condition $\Phi \to -G_N M/R_H $ for $R \to R_H$. Therefore, the boundary
conditions for the potential $\widetilde \Phi$ are:
%%%%%
\be{2.8}
\left.\frac{d\widetilde\Phi}{dR}\right|_{R=R_H}=0,\quad \widetilde\Phi(R_H)=-\frac{1}{2}\frac{\ddot a}{a}R_H^2-\frac{G_N M}{R_H}\, .
\ee
%%%%%%
%The average potential over the region inside the zero acceleration sphere is given by the following formula:
%%%%%%
%\be{2.9} \bar{\Phi}_{\Lambda CDM}=\frac{4\pi}{V_{H}}\int_{0}^{R_H}\Phi_{\Lambda CDM}(R)R^{2}dR,\quad V_H=\frac{4}{3}\pi R_{H}^{3}\, .\ee
%%%%%%
%
For the given above boundary conditions, the radius of the zero acceleration sphere is
%%%%%%
\be{2.10}
 R_{H}=\left[\frac{G_{N}M}{\ddot{a} /a}\right]^{1/3}\quad \Rightarrow \quad \widetilde\Phi(R_H)=-\frac32\frac{ G_N M}{R_H}\, .
\ee
%%%%%%
Obviously, the value of $R_H$ depends on time. On the other hand, Eq. \rf{2.10} results in a useful relation
%%%%%%
\be{2.10a}
\frac{\ddot{a}}{a} R_H^3 = G_N M\, .
\ee
%%%%%%
The ratio $\ddot a/a$ can be obtained from the second Friedmann equation. In the case of the $\Lambda$CDM model we have:
%%%%%
\be{2.11}
\frac{\ddot{a}}{a}=-\frac{\kappa \bar{\rho} c^{4}}{6a^{3}}+\frac{\Lambda
c^{2}}{3}=H_{0}^{2}\left(-\frac{1}{2}\Omega_{M}\frac{a_{0}^{3}}{a^{3}}+\Omega_{\Lambda}\right)>0\, ,
\ee
%%%%%%
where  $\kappa=8\pi G_N/c^4$ (with $c$ being the speed of light) and we took into account the late time acceleration of the Universe expansion. In accordance with
the conventional $\Lambda$CDM model, the Universe is supposed to be filled with the nonrelativistic matter (dust) characterized by the average rest mass density
$\bar\rho$ (being constant in the comoving reference frame) and the dark energy represented by the cosmological constant $\Lambda$. We also introduce  the
standard density parameters
%%%%%%
\be{2.12}
\Omega_M=\frac{\kappa\bar\rho c^4}{3H_0^2a_0^3},\quad \Omega_{\Lambda}=\frac{\Lambda c^2}{3H_0^2}\, ,
\ee
%%%%%%
where $a_0$ and $H_0$ are the current values of the scale factor and the Hubble parameter $H(t)=\dot a/a$, respectively. Therefore, from \rf{2.10} we get
%%%%%%
\be{2.13} R_H=\frac{(G_{N}M)^{1/3}}{\left[H_{0}^{2}\left(-\frac{1}{2}\Omega_{M}\frac{a_{0}^{3}}{a^{3}}+\Omega_{\Lambda}\right)\right]^{1/3}}\, .\ee
%%%%%

If instead of the discrete cosmology approach the Schwarzschild-de Sitter one is applied (see \citep{Chernin}), then in the above and following formulae we should
simply put $\Omega_M \equiv 0$, thus disregarding the contribution of matter (see also \citep{IJMPD} for the different argumentation). However, according to the
resent observations, matter contributes approximately 30\% into the total balance (dark energy/$\Lambda$-term gives another 70\%). Hence, dropping the
contribution of matter corresponds to a decrease in accuracy of 30 percent. At the present, we are entering the era of precision cosmology. Therefore, future
observations can reveal these discrepancies.

The mean velocity dispersion of galaxies $\sigma$ in clusters is one of their important observable parameters \citep{veldisp}. With respect to the Coma cluster,
the line-of-sight velocity dispersion is $954 \pm 50\, \mathrm{km}\, \mathrm{s}^{-1}$ \citep{Tully} (the authors of \citep{veldisp} obtained a slightly different
value of $\sigma \approx 1008\, \mathrm{km}\, \mathrm{s}^{-1}$).
%For example,  $\sigma \approx $ 700 km/s for the Virgo cluster \cite{KarNas}.
%With $\sigma\approx $ 1008 km/s for the Coma cluster (Abell %1656) \cite{veldisp} and $\sigma \approx $ 879 km/s for the Leo cluster (Abell 1367) \cite{veldisp}.
We can estimate $\sigma $ with the help of the virial theorem. Eq. \rf{2.4} can be written in the form
%%%%%
\be{2.23}
\dot {\bf{V}} = - \frac{\partial \widetilde\Phi}{\partial R}\frac{\bf{R}}{R}\, ,
\ee
%%%%%
%where $\Phi_i$ is either $\Phi_{\Lambda CDM}$ or $\Phi_{\Lambda}$.
Multiplying both sides of Eq. \rf{2.23} by ${\bf{R}}$, one gets:
%%%%%
\be{2.24}
\frac{d}{dt}({\bf V R})= V^2 - \frac{\partial \widetilde\Phi}{\partial R} R\, .
\ee
%%%%%
Because $V$ and $R$ are finite functions, averaging this equation over times which are much longer than characteristic periods of the system, we can eliminate the
left hand side of \rf{2.24}. Next, we average over the volume of the halo  and obtain
%%%%%%
\be{2.25} \overline{V^2} = \overline{\frac{\partial \widetilde\Phi}{\partial R} R}\, .\ee
%%%%%%
The mean velocity dispersion  $\sigma = (\, \overline {V^2}\, )^{1/2}$. If the number density of galaxies over the entire volume of the halo (i.e. for $R\in
[0,R_H]$) is constant, then
%%%%%%
\be{2.26}
\overline{V^2} = \frac{4\pi}{V_H}\int_0^{R_H} \frac{\partial \widetilde\Phi}{\partial R} R^3 dR\, ,\quad V_H = \frac43 \pi R_H^3\, .
\ee
%%%%%%
However, observations indicate that distribution of galaxies is abruptly bounded at some distance $R_{2t}$ which is considerably less than the size of the
cluster. Most galaxies are inside this distance from the center of mass of the Coma cluster \citep{Tully}. Therefore, the averaging over the entire volume of the
halo gives the wrong result. Unfortunately, we do not know the real 3D picture for vectors of velocities ${\bf V}$ of all galaxies in the Coma cluster. Hence, to
get the mean velocity dispersion we need to introduce some theoretical model. Let us suppose a toy model where all galaxies are concentrated inside a sphere of
the radius $R_{\mathrm{eff}}$ with the constant number density. There are no galaxies outside this sphere. For such picture,
%%%%%%
\be{2.27}
\overline{V^2} = \frac{4\pi}{V_{\mathrm{eff}}}\int_0^{R_{\mathrm{eff}}} \frac{\partial \widetilde\Phi}{\partial R} R^3 dR\, ,
\quad V_{\mathrm{eff}} = \frac43 \pi R_{\mathrm{eff}}^3\, .
\ee
%%%%%%

Now we want to take a definite DM halo profile and, first, calculate the gravitational potential $\widetilde\Phi$, second, obtain the mean velocity dispersion
from the formula \rf{2.27} as a function of $R_{\mathrm{eff}}$. As we know the observed value of $\sigma$ as well as the experimental values for the halo profile
parameters, we can estimate unknown $R_{\mathrm{eff}}$ and compare it with the observable $R_{2t}$. If these two values are close to each other, it will
demonstrate that our toy model determines more or less successfully the characteristic radius within which most of the galaxies are located. On the other hand, if
$R_{\mathrm{eff}}$ is close to the observable value, we can use this model to determine parameters of the considered DM halo profile. In the next section, we
demonstrate our approach on the example of the NFW profile.

\section{Navarro-Frenk-White profile in the Coma cluster. Theory vs observations and simulations}\label{sec:3}

At the present, the NFW profile \citep{NFW} is one of the most commonly used profiles to describe DM distribution in galaxies and clusters of galaxies. It reads
%%%%%
\be{3.1}
\rho_{\mathrm{ph}}(R)=\frac{4\rho_{s}}{\frac{R}{R_{s}}\left(1+\frac{R}{R_{s}}\right)^{2}}\, ,
\ee
%%%%%
where $\rho_s=\rho_{\mathrm{ph}}(R_s)$ and the scale radius $R_s$  are free parameters of the profile. Since DM gives the main contribution to the total mass of
the system, the mass $M_{*}$ within a sphere of the radius $R_{*}$ is
%%%%%
\be{3.2}
\ M_{*}\equiv M(R_{*}) = 16\pi
\rho_{s}R_{s}^{3}\left[\ln\left(1+\frac{R_{*}}{R_{s}}\right)-\frac{R_{*}}{R_s+R_{*}}\right]\, .
\ee
%%%%%%
Then, the total mass of the cluster is $M \equiv M(R_H)$. Very often DM profiles are characterized by the mass $M_{200}\equiv M(R_{200})$ where the radius
$R_{200}$ defines a sphere where the average DM density $\bar \rho_{200}$ equals 200 $\rho_{crit}$ with
%%%%%%
\be{3.3} \ \rho_{crit}=3H_0^2/(8\pi G_N)\approx1.88\,h^2\times 10^{-26} \, \mathrm{kg}\, \mathrm{m}^{-3} \ee
%%%%%%
being the critical density of the Universe today. According to the most recent Planck results \citep{Planck2}, $\Omega_M\approx0.3$ and
$\Omega_{\Lambda}\approx0.7$. Besides, $H_0\approx67.8$ $\mathrm{km}\,\mathrm{s}^{-1}\,\mathrm{Mpc}^{-1}$, therefore, $h\approx0.678$. Hence, by the definition
%%%%%%
\be{3.4}
M_{200}=\frac{4}{3}\pi R_{200}^3\bar\rho_{200}\, , \quad \bar\rho_{200}=200\rho_{crit} \, .
\ee
%%%%%%
It follows from Eq. \rf{2.13} that the total mass
%%%%%
\ba{3.5}
&{}&M = \frac{8\pi}{3}\rho_{crit}R_H^3\left(-\frac{1}{2}\Omega_M+\Omega_{\Lambda}\right) =\frac{M_{200}}{100}\left(\frac{\lambda_H}{c}\right)^3
\left(-\frac{1}{2}\Omega_M+\Omega_{\Lambda}\right)\, ,
\ea
%%%%%
where
%%%%%%
\be{3.6}
\lambda_H \equiv \frac{R_H}{R_s}\, , \quad c\equiv \frac{R_{200}}{R_s}\, .
\ee
%%%%%%
The parameter $c$ is usually called the concentration parameter. On the other hand, from Eq. \rf{3.2} we have
%%%%%%
\be{3.7}
M = M_{200}\frac{\ln(1+\lambda_H)-\frac{\lambda_H}{1+\lambda_H}}{\ln(1+c)-\frac{c}{1+c}}\, .
\ee
%%%%%%%
Equating \rf{3.5} and \rf{3.7}, we get the following useful relation:
%%%%%%%
\be{3.8}
\frac{1}{100}\left(-\frac{1}{2}\Omega_M+\Omega_{\Lambda}\right)\frac{1}{c^3}\left[\ln(1+c)-\frac{c}{1+c}\right] =
\frac{1}{\lambda_H^3}\left[\ln(1+\lambda_H)-\frac{\lambda_H}{1+\lambda_H}\right]\, .
\ee
%%%%%%
The ranges of parameters for the Coma cluster halo according to the observations are\footnote{It is worth noting that if we calculate $M_{200}^{(min,max)}$ via
the formula \rf{3.4} with $R_{200}^{(min,max)}$ from \rf{3.9}, then we get slightly different values: $M_{200}^{(min)}= 1.29 h^{-1}\times10^{15}M_{\odot}$ and
$M_{200}^{(max)}= 2.47 h^{-1}\times10^{15}M_{\odot}$.} \citep{Kubo}:
%%%%%%
\ba{3.9}
&{}& R_{200}^{(min)} = 1.77\,h^{-1} \, \mathrm{Mpc} = 2.61\, \mathrm{Mpc}\, ,\quad R_{200}^{(max)} = 2.20\,h^{-1}\, \mathrm{Mpc} = 3.24\, \mathrm{Mpc}\, ,\nn \\
&{}& M_{200}^{(min)}= 1.32 h^{-1}\times10^{15}M_{\odot}\, ,\quad M_{200}^{(max)}= 2.53 h^{-1}\times10^{15}M_{\odot}\, ,\nn\\
&{}&c_{min}=2\, ,\quad c_{max} =17 \, .\ea
%%%%%%
In these formulae, min and max values follow from the error bars in the analysis for the corresponding parameters in \citep{Kubo}.

Let us turn now to the gravitational potential and the velocity dispersion. To get the expression for the gravitational potential $\widetilde \Phi$ defined in
\rf{2.5}, we should first solve the Poisson equation \rf{2.2} and, then, use the boundary conditions \rf{2.8}. As a result, we obtain:
%%%%%%%
\ba{3.10}
&{}& \widetilde\Phi(R)=-\frac{\ddot a}{a}\left(\frac{R^2}{2}-R_{H}^{2}+\frac{R_{H}^{3}}{R}\right) +
16\pi G_{N}\rho_{s}\frac{R_{s}^{3}}{R_{s}+R_{H}}\left(1-\frac{R_{H}}{R}\right)\nn\\
&-&16\pi G_{N}\rho_{s}\frac{R_{s}^{3}}{R}\ln\frac{R_{s}+R}{R_{s}+R_{H}}-\frac{G_N M}{R_H} =
-\frac{G_N M}{R_{H}}\left\{\frac{R^2}{2R_{H}^2}+\frac{R_H}{R}\right.\nn\\
&{}&\left.+\frac{R_H}{R}\frac{1}{\ln\left(1+\lambda_H\right)- \frac{\lambda_H}{1+\lambda_H}}\times\left[
\frac{\lambda_H -R/R_s}{1+\lambda_H}+\ln\frac{1+R/R_s}{1+\lambda_H}\right]\right\}\, ,
\ea
%%%%%%
where for $\ddot a/a$ we used Eq. \rf{2.10a} and $\rho_s$ is expressed with the help of Eq. \rf{3.2} for $R_{*}=R_H$. Now, we can calculate the velocity
dispersion with the help of Eq. \rf{2.27}. Instead of $\left[\overline{V^2}\right]^{1/2}$, it makes sense to consider the dimensionless quantity $\tilde\sigma$:
%%%%%%%
\ba{3.11} &{}& \widetilde\sigma =\left[\overline{V^2}\frac{R_{H}}{G_N M}\right]^{1/2} =
\left(\frac{\lambda_{H}}{\lambda_{\mathrm{eff}}}\right)^{1/2}\left[-\frac{3}{5}\left(\frac{\lambda_{\mathrm{eff}}}{\lambda_{H}}\right)^{3}
+\frac{3}{2}+\frac{1}{\ln\left(1+\lambda_{H}\right)- \frac{\lambda_{H}}{1+\lambda_{H}}}\right. \nn\\
&{}&\left.\times\left[ \frac{3}{2}\frac{\lambda_{H}}{1+\lambda_{H}} + \frac{9}{2}\frac{1}{\lambda_{\mathrm{eff}}} -\frac{9}{4}\right. -\left.
\frac{9}{2}\frac{1}{\lambda_{\mathrm{eff}}^2}\ln\left(1+\lambda_{\mathrm{eff}}\right) + \frac{3}{2}\ln\frac{1+\lambda_{\mathrm{eff}}}{1+\lambda_{H}}\right]
\right]^{1/2}\, ,\ea
%%%%%%%
where
%%%%%%
\be{3.12}
\lambda_{\mathrm{eff}} \equiv \frac{R_{\mathrm{eff}}}{R_s}\, .
\ee
%%%%%%
On the other hand, from the definition of $\widetilde\sigma$ we get
%%%%%
\be{3.13}
\widetilde\sigma =\left[\overline{V^2}\frac{R_{H}}{G_NM}\right]^{1/2}= \left[\overline{V^2}\right]^{1/2} \times\left[\frac{3}{8\pi}
\frac{c^2}{G_N\lambda_{H}^2R_{200}^2\rho_{crit}} \left(-\frac{1}{2}\Omega_M+\Omega_{\Lambda}\right)^{-1}\right]^{1/2}\, ,
\ee
%%%%%
where $R_H$ is expressed as $R_H=R_{200}\lambda_H/c$ and the total mass $M$ is given by Eq. \rf{3.5} with $M_{200}$ from \rf{3.4}.

According to the observations \citep{Tully}, the line-of-sight velocity dispersion for the Coma cluster is $954\pm50\, \mathrm{km}\, \mathrm{s}^{-1}$ (see Fig.~1
in \citep{Tully}, where the galaxy sample used for the estimation of $\sigma$ and their spatial extension are discussed). To get the 3D velocity dispersion we
should multiply this value by $\sqrt{3}$. Therefore, the minimal and maximal 3D velocity dispersions are: $\left[\overline{(V^{(min)})^2}\right]^{1/2}=1566\,
\mathrm{km}\, \mathrm{s}^{-1}$ and $\left[\overline{(V^{(max)})^2}\right]^{1/2}=1739\, \mathrm{km}\, \mathrm{s}^{-1}$, respectively. Therefore, the minimal and
maximal values of $\widetilde\sigma$ are:
%%%%%%
\be{3.14} \widetilde\sigma^{(min)} = \left[\overline{(V^{(min)})^2}\right]^{1/2}
 \times\left[\frac{3}{8\pi}\frac{c^2}{G_N\lambda_{H}^2(R_{200}^{(max)})^2\rho_{crit}} \left(-\frac{1}{2}\Omega_M+\Omega_{\Lambda}\right)^{-1}\right]^{1/2}\, ,\ee
\be{3.15} \widetilde\sigma^{(max)} = \left[\overline{(V^{(max)})^2}\right]^{1/2}
 \times\left[\frac{3}{8\pi}\frac{c^2}{G_N\lambda_{H}^2(R_{200}^{(min)})^2\rho_{crit}} \left(-\frac{1}{2}\Omega_M+\Omega_{\Lambda}\right)^{-1}\right]^{1/2}\, ,\ee
%%%%%%%%
where $R_{200}^{(min,max)}$ should be taken from \rf{3.9}.

According to Eq. \rf{3.9}, for the Coma cluster the concentration parameter $c\in [2,17]$. Our task is to determine the values of $R_{\mathrm{eff}}$ as a function
of $c$ from this region. To perform it, we can start from $c=2$ and then define $R_{\mathrm{eff}}$ for each value of $c$ with the step, e.g., $\triangle{c} =1$ up
to $c=17$. It is convenient to present the obtained values of $R_{\mathrm{eff}}$ in the corresponding table $R_{\mathrm{eff}}$ vs $c$. To demonstrate this
procedure, let us consider as an example the case $c=4$. For a given value of $c$, the parameter $\lambda_H$ can be found from Eq. \rf{3.8}. In the considered
example this is $\lambda_H = 33.28$. For this value of $\lambda_H$, the graph of $\widetilde\sigma$ (see Eq. \rf{3.11}) as a function of $\lambda_{\mathrm{eff}}$
is drawn in Fig. 1. From Eqs. \rf{3.14} and \rf{3.15}, we can also get for $c=4$: $\widetilde\sigma^{(min)} = 1.15$ (lower horizontal line in Fig.1) and
$\widetilde\sigma^{(max)} = 1.60$ (upper horizontal line in Fig.1). It can be easily seen from this picture that for each value of $\widetilde\sigma^{(min,max)}$
we have two roots. They are $\lambda_{\mathrm{eff},1}=0.39$ and $\lambda_{\mathrm{eff},2}=21.31$ for $\widetilde\sigma^{(min)}$, $\lambda_{\mathrm{eff},3}=1.58$
and $\lambda_{\mathrm{eff},4}=5.68$ for $\widetilde\sigma^{(max)}$. Since $\widetilde\sigma^{(min)}\, \left(\widetilde\sigma^{(max)}\right)$ was obtained with the
help of $R_{200}^{(max)}\, \left(R_{200}^{(min)}\right)$, the corresponding values for $R_H=R_{200}\lambda_H/c$ and $R_s=R_{200}/c$ are $R_H^{(max)}=26.99$ Mpc
and $R_s^{(max)}=0.81$ Mpc ($R_H^{(min)}=21.72$ Mpc and $R_s^{(min)}=0.65$ Mpc). Therefore, from the relation $R_{\mathrm{eff}}=\lambda_{\mathrm{eff}} R_s$ we
obtain four marginal  values: $R_{\mathrm{eff},1}^{(max)}=0.32$ Mpc and $R_{\mathrm{eff},2}^{(max)}=17.28$ Mpc (for $R_s^{(max)}$ and $\lambda_{\mathrm{eff}1,2}$,
correspondingly) and $R_{\mathrm{eff},3}^{(min)}=1.03$ Mpc and $R_{\mathrm{eff},4}^{(min)}=3.71$ Mpc (for $R_s^{(min)}$ and $\lambda_{\mathrm{eff}3,4}$,
correspondingly).

These calculations demonstrate that only $R_{\mathrm{eff},4}^{(min)}$ is close to the observable value $R_{2t}\sim 3$ Mpc \citep{Tully}. Calculations for other
values of the concentration parameter $c$ from the considered region lead to the same conclusion. Hence, only the marginal value  $R_{\mathrm{eff},4}^{(min)}$
corresponds to the observations. We summarize the results of our calculations in Table 1.

\

%%%%%%%%%%%%%%\prod
\begin{figure}[h]
\centerline{\includegraphics[width=3.0in,height=2in]{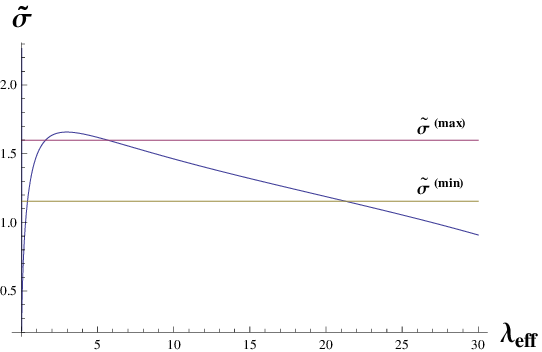}}
\caption {This plot shows the behavior of the dimensionless velocity dispersion $\widetilde \sigma$ as a function of $\lambda_{\mathrm{eff}}$ for the concentration
parameter $c=4$ (curved line). Upper and lower horizontal lines correspond to the maximal and minimal values of $\widetilde \sigma$ obtained with the help of the
observable data.\label{c4sec:1}}
\end{figure}
%%%%%%%%%%%%%%%%%%%%%%%%%%%%

\

\begin{table}[h]
  \begin{center}
   \caption{The values of the root $R_{\mathrm{eff},4}^{(min)}$ for different values of the concentration parameter $c$.}
    \setlength{\tabcolsep}{11pt}
    \begin{tabularx}{\textwidth}{ccccccccc}
    \hline
      $R^{(min)}_{\mathrm{eff},4}$ & $4.17$ & $3.70$ & $3.71$ & $3.72$ & $3.74$ & $3.75$ & $3.76$ & $3.77$ \\
      c & 2 & 3 & 4 & 5 & 6 & 7 & 8 & 9 \\
      \renewcommand{\arraystretch}{1.1}
    \end{tabularx}
    \begin{tabularx}{\textwidth}{ccccccccc}
      $R^{(min)}_{\mathrm{eff},4}$ & $3.78$ & $3.78$ &$3.79$ & $3.80$ & $3.80$ & $3.80$ & $3.81$ & $3.81$ \\
      c & 10 & 11 & 12 & 13 & 14 & 15 & 16 & 17 \\
    \hline
    \end{tabularx}
   \label{tab:table1}
  \end{center}
\end{table}

The table shows that the closest with respect to the observations  results take place for the values $c=3\div 4$ where $R_{\mathrm{eff},4}^{(min)}$ has a minimum.

It is worth noting that this concentration parameter is very close to the one obtained from the fitting of the NFW model. This fit is based on a series of DM
cosmological simulations within the $\Lambda$CDM model \citep{DuttMac,Sch}. For example, if we take $M_{200}^{(min)}= 1.29 h^{-1}\times10^{15}M_{\odot}$ (see
footnote 1), then we get from the equations for the concentration-mass relations (see Eq. (24) in \citep{Sch} or equivalently Eq. (8) in \citep{DuttMac}) that
$c\approx 3.89$.

Therefore, the comparison of our toy model with the observable data and simulations leads to the following preferable NFW parameters for the Coma cluster:
$R_{200} \approx 1.77\,h^{-1} \, \mathrm{Mpc} = 2.61\, \mathrm{Mpc}$ and $c=3\div 4$. For these values of $c$ and $M_{200}= 1.29 h^{-1}\times10^{15}M_{\odot}$,
the total mass of the Coma cluster in the case of the NFW profile is $M = (4.08\div 4.74)\,h^{-1}\times 10^{15} M_{\odot}$ (see Eq. \rf{3.7}). We also found that
in this cluster the most of galaxies are concentrated inside a sphere of the effective radius $R_{\mathrm{eff}}\sim 3.7$ Mpc and the line-of-sight velocity
dispersion is $1004\, \mathrm{km}\, \mathrm{s}^{-1}$. This value is very close to $1008\, \mathrm{km}\, \mathrm{s}^{-1}$ obtained from the observations for the
Coma cluster (Abell 1656) in \citep{veldisp}.

\section{Conclusion}\label{sec:4}

In short, the main results of our paper are the following. First, within the discrete cosmology approach, we proposed the method to obtain the gravitational
potential for any spherically distributed profiles. For this purpose, we defined the corresponding boundary conditions on the surface of the zero acceleration.
The notion of this surface was defined in our previous papers \citep{EZcosm1,EZcosm2,EKZ2}. It enabled us to introduce the concept of the cluster size. We
naturally supposed that the edge/cutoff of a DM halo coincides with the surface of the zero acceleration, and the mass of a halo is defined by integration of the
rest mass density profile up to the radius of the zero acceleration sphere. Second, since we know now the expression for the gravitational potential, the virial
theorem enables us to get the formula for the mean velocity dispersion. We derived this formula using the observations \citep{Tully} that distribution of galaxies
in clusters is abruptly bounded at some distance which we denoted by $R_{\mathrm{eff}}$. This formula can be applied for any cluster with spherical mass
distribution. This relationship can be used as an additional constraint on the parameters of the DM profiles. This is one of the main points of our paper.
Moreover, we can consider any spherical DM profiles within our approach but not only the NFW one. Third, we applied this approach to the Coma cluster to estimate
the parameters of the NFW density profile for this cluster.

In more detail, in this paper we have investigated the distribution of visible and dark matter in the Coma cluster. In the late Universe, such clusters are
gravitationally bound systems and can be considered as isolated objects. To study them, we can apply the discrete cosmology (mechanical) approach
\citep{EZcosm1,EZcosm2,XXX}. In this approach, the motion of a test body in the vicinity of an inhomogeneity is defined by two factors: the gravitating attraction
of the inhomogeneity and the accelerated cosmological expansion of the Universe. These two competitive mechanisms define a region where the acceleration of a test
body is equal to zero: the zero acceleration surface. In the case of spherical distribution of matter this surface is a sphere of the radius $R_H$. Inside this
sphere, the gravitational attraction plays the main role while at the larger distances the cosmological expansion prevails. This is the region of the Hubble flows
formation. It is well known that the DM halo gives the main contribution to the total mass of a cluster. Hence, the gravitational potential of a cluster is mainly
defined by the DM distribution in it. There are a number of halo profiles proposed to describe the DM distribution in galaxies and clusters of galaxies. It is
natural to define the edge of the DM halo at the radius of zero acceleration. Because the halo of DM dominates in clusters of galaxies, the visible matter in the
form of galaxies is considered as test bodies in the gravitational field of the halo.

In our paper, we have investigated the Poisson equation for the gravitational potential of a spherically symmetric halo profile. We have defined the proper
boundary conditions at $R_H$. Then, with the help of the virial theorem, we have got the formula for the mean velocity dispersion for the visible matter
(galaxies) inside this halo. Observations show \citep{Tully} that galaxies inside clusters are abruptly bounded at some distance $R_{2t}$ which is considerably
less that the size of the cluster. Most galaxies are inside this distance from the center of mass of the Coma cluster \citep{Tully}. Since we do not know the real
3D velocities of galaxies,  we have proposed a toy model where all galaxies are concentrated inside a sphere of an effective radius $R_{\mathrm{eff}}$ with the
constant number density. This has enabled us to get the expressions for the mean velocity dispersion for an arbitrary spherically symmetric halo profile.

To perform specific calculations, we need to suggest a definite form for the DM density profile. The NFW profile \citep{NFW} is one of the most commonly used. We
have used this profile in the case of the Coma cluster. First, we have got the formula for the gravitational potential in this profile. Then, we have obtained the
mean velocity dispersion as a function of $R_{\mathrm{eff}}$. We have shown that the calculated value of $R_{\mathrm{eff}}$ can be rather close to the observable
$R_{2t}$ within the observation accuracy of the NFW parameters. Moreover, the comparison of our toy model with the observable data and simulations leads to the
following preferable NFW parameters for the Coma cluster: $R_{200} \approx 1.77\times h^{-1} \, \mathrm{Mpc} = 2.61\, \mathrm{Mpc}$, $c=3\div 4$ and $M_{200}=
1.29 h^{-1}\times10^{15}M_{\odot}$. We have also found that in this cluster most of galaxies are concentrated inside a sphere of the effective radius
$R_{\mathrm{eff}}\sim 3.7$ Mpc, and the line-of-sight velocity dispersion is $1004\, \mathrm{km}\, \mathrm{s}^{-1}$.

%%%%%%%%%%%%%%%%%%%%%%%%%%%%%%%%%%%%%%%%%%%%%%%%%%%%%%%%%%%%%%%%%%%%%%%%%%%
\section*{Acknowledgements}
We would like to thank V.E. Karachentseva for very useful comments. The work of R.~Brilenkov was partially supported by the EMJMD Student Scholarship from the
Erasmus\,+\,: Erasmus Mundus Joint Master Degree programme AstroMundus in Astrophysics. The work of M.~Eingorn was partially supported by NSF CREST award
HRD-1345219 and NASA grant NNX09AV07A.

%%%%%%%%%%%%%%%%%%%%%%%%%%%%%%%%%%%%%%%%%%%%%%%%%%%%%%%%%%%%%%%%%%%%%%%%%%


\begin{thebibliography}{99}

\bibitem[\protect\citeauthoryear{{Ade} et al.}{2014}]{Planck}
% authors
Ade P.A.R. et al.:
% article title
Planck 2013 results. XVI. Cosmological parameters.
% journal title, volume, first page, year of publication
AAp\ \textbf{517}, A16 (2014)

\bibitem[\protect\citeauthoryear{{Ade} et al.}{2016}]{Planck2}
% authors
Ade P.A.R. et al.:
% article title
Planck 2015 results. XIII. Cosmological parameters. AAp\ \textbf{594}, A13 (2016)

\bibitem[\protect\citeauthoryear{{Binggeli}, {Sandage}, {Tamman}}{1985}]{BinSanTamm1}
% authors
Binggeli B., Sandage A., Tamman G.A.:
% article title
Studies of the Virgo cluster. II. A catalog of 2096 galaxies in the Virgo cluster area.
% journal title, volume, first page, year of publication
AJ\ \textbf{90}, 1681 (1985)

\bibitem[\protect\citeauthoryear{{Binggeli}, {Sandage}, {Tamman}}{1987}]{BinSanTamm2}
% authors
Binggeli B., Sandage A., Tamman G.A.:
% article title
Studies of the Virgo cluster. VI. Morphological and kinematical structure of the Virgo cluster.
% journal title, volume, first page, year of publication
AJ\ \textbf{94}, 251 (1987)

\bibitem[\protect\citeauthoryear{{Bisnovatyi-Kogan}, {Chernin}}{2011}]{Chernin}
% authors
Bisnovatyi-Kogan G.S., Chernin A.D.:
% article title
Dark energy and key physical parameters of clusters of galaxies.
% journal title, volume, first page, year of publication
ApSS\ \textbf{338}, 337 (2012)

\bibitem[\protect\citeauthoryear{{Boselli} et al.}{2014}]{Boselli}
% authors
Boselli A.:
% article title
The GALEX Ultraviolet Virgo Cluster Survey (GUViCS). IV: The role of the cluster environment on galaxy evolution.
% journal title, volume, first page, year of publication
AAp\ \textbf{570}, A69 (2014)

\bibitem[\protect\citeauthoryear{{Boyarsky} et al.}{2009}]{Boyar1}
% authors
Boyarsky A. et al.:
% article title
New evidence for dark matter.
% journal title, volume, first page, year of publication
arXiv:0911.1774 [astro-ph], (2009)

\bibitem[\protect\citeauthoryear{{Boyarsky} et al.}{2010}]{Boyar2}
% authors
Boyarsky A. et al.:
% article title
Universal properties of Dark Matter halos.
% journal title, volume, first page, year of publication
PRL\ \textbf{104}, 191301 (2010)



\bibitem[\protect\citeauthoryear{{Brilenkov}, {Eingorn}}{2017}]{BrilEin}
% authors
Brilenkov R., Eingorn M.:
% article title
Second-order Cosmological Perturbations Engendered by Point-like Masses.
% journal title, volume, first page, year of publication
ApJ\ \textbf{845}, 153 (2017) %arXiv:1703.10282 [gr-qc]



\bibitem[\protect\citeauthoryear{{Chemin}, {de Blok}, {Mamon}}{2011}]{Chemin}
% authors
Chemin L., de Blok W.J.G., Mamon G.A.:
% article title
Improved Modeling of the Mass Distribution of Disk Galaxies by the Einasto Halo Model.
% journal title, volume, first page, year of publication
AJ\ \textbf{142}, 109 (2011)

\bibitem[\protect\citeauthoryear{{Chernin}, {Teerikorpi}, {Baryshev}}{2003}]{Chernin1}
% authors
Chernin A.D., Teerikorpi P., Baryshev Yu.:
% article title
Why is the Hubble flow so quiet?
% journal title, volume, first page, year of publication
Adv. Space Res.\ \textbf{31}, 459 (2003)

\bibitem[\protect\citeauthoryear{{Diemand}, {Moore}}{2011}]{minireview1}
% authors
Diemand J., Moore B.:
% article title
The structure and evolution of cold dark matter halos.
% journal title, volume, first page, year of publication
Adv. Sci. Lett.\ \textbf{4}, 297 (2011)

\bibitem[\protect\citeauthoryear{{Dutton}, {Maccio}}{2014}]{DuttMac}
% authors
Dutton A.A., Maccio A.V.:
% article title
Cold dark matter haloes in the Planck era: evolution of structural parameters for Einasto and NFW profiles.
% journal title, volume, first page, year of publication
MNRAS\ \textbf{441}, 3359 (2014)

\bibitem[\protect\citeauthoryear{{Eingorn}, {Zhuk}}{2012}]{EZcosm1}
% authors
Eingorn M., Zhuk A.:
% article title
Hubble flows and gravitational potentials in observable Universe.
% journal title, volume, first page, year of publication
JCAP\ \textbf{09}, 026 (2012)

\bibitem[\protect\citeauthoryear{{Eingorn}, {Kudinova}, {Zhuk}}{2013}]{EKZ2}
% authors
Eingorn M., Kudinova A., Zhuk A.:
% article title
Dynamics of astrophysical objects against the cosmological background.
% journal title, volume, first page, year of publication
JCAP\ \textbf{04}, 010 (2013)

\bibitem[\protect\citeauthoryear{{Eingorn}, {Zhuk}}{2014}]{EZcosm2}
% authors
Eingorn M., Zhuk A.:
% article title
Remarks on mechanical approach to observable Universe.
% journal title, volume, first page, year of publication
JCAP\ \textbf{05}, 024 (2014)



\bibitem[\protect\citeauthoryear{{Eingorn}, {Brilenkov}}{2015}]{Brilenkov}
% authors
Eingorn M., Brilenkov R.:
% article title
Perfect fluids with $\omega=\mathrm{const}$ as sources of scalar cosmological perturbations.
% journal title, volume, first page, year of publication
Physics of the Dark Universe\ \textbf{17}, 63 (2017) %arXiv:1509.08181 [gr-qc]



\bibitem[\protect\citeauthoryear{{Eingorn}}{2016}]{XXX}
% authors
Eingorn M.:
% article title
First-order Cosmological Perturbations Engendered by Point-like Masses.
% journal title, volume, first page, year of publication
ApJ\ \textbf{825}, 84 (2016)



\bibitem[\protect\citeauthoryear{{Eingorn}, {Kiefer}, {Zhuk}}{2016}]{Kiefer}
% authors
Eingorn M., Kiefer C., Zhuk A.:
% article title
Scalar and vector perturbations in a universe with discrete and continuous matter sources.
% journal title, volume, first page, year of publication
JCAP\ \textbf{09}, 032 (2016)



\bibitem[\protect\citeauthoryear{{Eingorn}}{2017}]{IJMPD}
% authors
Eingorn M.:
% article title
Cosmological law of universal gravitation.
% journal title, volume, first page, year of publication
Int. J. Mod. Phys. D\ \textbf{26}, 1750121 (2017)



\bibitem[\protect\citeauthoryear{{Eisenstein} et al.}{2005}]{Eisenstein}
% authors
Eisenstein D.J. et al.:
% article title
Detection of the Baryon Acoustic Peak in the Large-Scale Correlation Function of SDSS Luminous Red Galaxies.
% journal title, volume, first page, year of publication
ApJ\ \textbf{633}, 560 (2005)

\bibitem[\protect\citeauthoryear{{Hinshaw} et al.}{2013}]{Hinshaw}
% authors
Hinshaw G. et al.:
% article title
Nine-Year Wilkinson Microwave Anisotropy Probe (WMAP) Observations: Cosmological Parameter Results.
% journal title, volume, first page, year of publication
ApJ\ \textbf{208}, 19 (2013)

\bibitem[\protect\citeauthoryear{{Karachentsev} et al.}{2002}]{Kar2002}
% authors
Karachentsev I.D. et al.:
% article title
The very local Hubble flow.
% journal title, volume, first page, year of publication
AAp\ \textbf{389}, 812 (2002)

\bibitem[\protect\citeauthoryear{{Karachentsev}, {Kashibadze}, {Makarov}}{2009}]{Kar2008}
% authors
Karachentsev I.D., O.G. Kashibadze, Makarov D.I., Tully R.B.:
% article title
The Hubble flow around the Local Group.
% journal title, volume, first page, year of publication
MNRAS\ \textbf{393}, 1265 (2009)

\bibitem[\protect\citeauthoryear{{Karachentsev}}{2012}]{Kar2012}
% authors
Karachentsev I.D.:
% article title
Missing dark matter in the local Universe.
% journal title, volume, first page, year of publication
MNRAS\ \textbf{67}, 123 (2009)

\bibitem[\protect\citeauthoryear{{Karachentsev}, {Nasonova}}{2010}]{KarNas}
% authors
Karachentsev I.D., Nasonova O.G.:
% article title
The observed infall of galaxies towards the Virgo cluster.
% journal title, volume, first page, year of publication
MNRAS\ \textbf{405}, 1075 (2010)

\bibitem[\protect\citeauthoryear{{Kubo}}{2007}]{Kubo}
% authors
Kubo J.M.:
% article title
The Mass Of The Coma Cluster From Weak Lensing In The Sloan Digital Sky Survey.
% journal title, volume, first page, year of publication
AJ\ \textbf{671}, 1466 (2007)

\bibitem[\protect\citeauthoryear{{Lau} et al.}{2015}]{selfsim}
% authors
Lau E.T. et al.:
% article title
Mass Accretion and its Effects on the Self-Similarity of Gas Profiles in the Outskirts of Galaxy
Clusters.
% journal title, volume, first page, year of publication
ApJ\ \textbf{806}, 68 (2015)

\bibitem[\protect\citeauthoryear{{Lukovic}, {Cabella}, {Vittorio}}{2014}]{minireview2}
% authors
Lukovic V., Cabella P., Vittorio N.:
% article title
Dark Matter in Cosmology.
% journal title, volume, first page, year of publication
Int. J. Mod. Phys. A\ \textbf{29}, 1443001 (2014)

\bibitem[\protect\citeauthoryear{{Melchiorri}, {Pagano}, {Pandolfi}}{2007}]{accel}
% authors
Melchiorri A., Pagano L., Pandolfi S.:
% article title
When Did Cosmic Acceleration Start?
% journal title, volume, first page, year of publication
PRD\ \textbf{76}, 041301 (2007)

\bibitem[\protect\citeauthoryear{{Navarro}, {Frenk}, {White}}{1996}]{NFW}
% authors
Navarro J.F., Frenk C.S., White. S.D.M.:
% article title
The structure of Cold Dark Matter Halos.
% journal title, volume, first page, year of publication
AJ\ \textbf{462}, 563 (1996)

\bibitem[\protect\citeauthoryear{{Novosyadlyj} et al.}{2006}]{ourbook}
% authors and year
Novosyadlyj B. et al.:
% book title
Dark Energy: Observational Evidence and Theoretical Models (1st volume of three-volume book ``Dark energy and dark matter in the Universe", ed. V. Shulga),
Akademperiodyka, Kiev (2013)

\bibitem[\protect\citeauthoryear{{Okabe} et al.}{2010}]{Okabe}
% authors
Okabe N. et al.:
% article title
LoCuSS: Subaru Weak Lensing Study of 30 Galaxy Clusters.
% journal title, volume, first page, year of publication
PASJ\ \textbf{62}, 811 (2010)

\bibitem[\protect\citeauthoryear{{Peebles}}{1980}]{Peebles}
% authors
Peebles P.J.E.:
% article title
The large-scale structure of the Universe.
% journal title, volume, first page, year of publication
Princeton University Press, Princeton (1980)

\bibitem[\protect\citeauthoryear{{Perlmutter} et
    al.}{1999}]{Perlmutter}
% authors
Perlmutter S. et al.:
% article title
Measurements of Omega and Lambda from 42 High-Redshift Supernovae.
% journal title, volume, first page, year of publication
ApJ\ \textbf{517}, 565 (1999)

\bibitem[\protect\citeauthoryear{{Riess} et al.}{1998}]{Ries}
% authors
Riess A.G. et al.:
% article title
Observational Evidence from Supernovae for an Accelerating Universe and a Cosmological Constant.
% journal title, volume, first page, year of publication
AJ\ \textbf{116}, 1009 (1998)

\bibitem[\protect\citeauthoryear{{Saburova}, {Popolo}}{2014}]{SabPop}
% authors
Saburova A., Del Popolo A.:
% article title
On the surface density of dark matter haloes.
% journal title, volume, first page, year of publication
MNRAS\ \textbf{445}, 3512 (2014)

\bibitem[\protect\citeauthoryear{{Sasaki} et al.}{2015}]{subhalos}
% authors
Sasaki T. et al.:
% article title
Suzaku observations of subhalos in the Coma cluster.
% journal title, volume, first page, year of publication
arXiv:1504.03044 [astro-ph] (2015)

\bibitem[\protect\citeauthoryear{{Schaller} et al.}{2015}]{Sch}
% authors
Schaller M. et al.:
% article title
The masses and density profiles of halos in a LCDM galaxy formation simulation.
% journal title, volume, first page, year of publication
MNRAS\ \textbf{451}, 1247 (2015)

\bibitem[\protect\citeauthoryear{{Springel}}{2005}]{gadget2}
% authors
Springel V.:
% article title
The cosmological simulation code GADGET-2.
% journal title, volume, first page, year of publication
MNRAS\ \textbf{364}, 1105 (2005)

\bibitem[\protect\citeauthoryear{{Struble}, {Rood}}{1999}]{veldisp}
% authors
Struble M.F., Rood. H.J.:
% article title
A compilation of redshifts and velocity dispersions for ACO clusters.
% journal title, volume, first page, year of publication
ApJ\ \textbf{125}, 35 (1999)

\bibitem[\protect\citeauthoryear{{Tully}}{2015}]{Tully}
% authors
Tully R.B.:
% article title
Galaxy Groups.
% journal title, volume, first page, year of publication
AJ\ \textbf{149}, 54 (2015)

\bibitem[\protect\citeauthoryear{{Vikhlinin} et al.}{2006}]{clustermass}
% authors
Vikhlinin A. et al.:
% article title
Chandra sample of nearby relaxed galaxy clusters: mass, gas
fraction, and mass-temperature relation.
% journal title, volume, first page, year of publication
ApJ\ \textbf{640}, 691 (2006)

\end{thebibliography}
\end{document}